\documentclass[pra,twocolumn,showpacs,floatfix]{revtex4}
\usepackage{graphicx}
\begin{document}

\title{Rotating Bose-Einstein condensates with a finite number of atoms confined in a ring potential: 
Spontaneous symmetry breaking, beyond the mean-field approximation}

\author{A. Roussou$^1$, J. Smyrnakis$^2$, M. Magiropoulos$^2$, Nikolaos K. Efremidis$^1$, and G. M. Kavoulakis$^2$}
\affiliation{$^1$Department of Applied Mathematics, University of Crete, GR-71004, Heraklion, Greece \\
$^2$Technological Education Institute of Crete, P.O. Box 1939, GR-71004, Heraklion, Greece}

\date{\today}

\begin{abstract}

Motivated by recent experiments on Bose-Einstein condensed atoms which rotate in annular/toroidal traps 
we study the effect of the finiteness of the atom number $N$ on the states of lowest energy for a fixed 
expectation value of the angular momentum, under periodic boundary conditions. To attack this problem, 
we develop a general strategy, considering a linear superposition of the eigenstates of the many-body 
Hamiltonian, with amplitudes that we extract from the mean field approximation. This many-body state breaks 
the symmetry of the Hamiltonian, it has the same energy to leading order in $N$ as the mean-field state and 
the corresponding eigenstate of the Hamiltonian, however it has a lower energy to subleading order in $N$ 
and thus it is energetically favorable.

\end{abstract}
 \maketitle

\section{Introduction}

Several recent experiments in the field of cold atomic gases have managed to rotate, and even create 
persistent currents in clouds of Bose-Einstein condensed atoms which are confined in annular/toroidal 
traps \cite{Kurn, Olson, Phillips1, Foot, GK, Moulder, Ryu, Zoran}. Furthermore, the phenomenon of 
hysteresis has also been observed in an annular trap \cite{hysteresis}. Thus, a question which arises 
naturally from these experiments is what is the state of lowest energy of the atoms for a fixed 
expectation value of their angular momentum. 

Within the mean-field, Gross-Pitaevskii, approximation the answer to this question is given by the 
well-known solitary-wave solutions \cite{yrsol, Carr}, which have been investigated thoroughly 
\cite{solrefs}. When a trapping potential is present -- as in the case of cold atomic gases -- 
interesting phenomena arise. The easiest problem is that of an infinite system in the longitudinal 
direction, with a very tight trapping potential in the transverse direction. In this case the transverse 
degrees of freedom are frozen and the problem essentially reduces to that of an infinite line. As the 
transverse trapping potential becomes less tight and/or the interaction strength increases, the transverse 
degrees of freedom start to play a role, and as a result deviations from the standard quadratic nonlinear 
Schr\"odinger equation \cite{JKP} arise. 

Another interesting possibility is that of a ring-like trap, where one has to impose periodic boundary 
conditions. In this case the solutions are given by Jacobi elliptic functions \cite{Carr2, solfin}. 
Depending on the ratio between the coherence length and the periphery of the ring the density of these
solutions is either sinusoidal, or exponentially localized \cite{solfin}. In an infinite system the 
``dark" solution (i.e., the one with a density notch) is also static. On the other hand -- as a result 
of the periodic boundary conditions -- in a ring of a finite length the dark solitary wave has a finite 
velocity, while the static solitary-wave solution is ``grey" (i.e., the lowest value of the density is 
nonzero), however no solution exists which is both dark and static \cite{solfin}.

The situation becomes much less clear when one wants to introduce correlations and go beyond the mean-field 
approximation, which is actually the subject of the present study. More specifically, in what follows below, 
we investigate the effect of the finiteness of the atom number $N$ (assumed to be of order unity) on the state 
of lowest energy, for a fixed expectation value of the angular momentum. We stress that this question is not 
only interesting theoretically, but also it is experimentally relevant, since recent experiments have managed 
to trap and detect very small numbers of atoms, see, e.g., Ref.\,\cite{SJ}. We should also mention that other
studies \cite{relationship, tgw, rela2} have investigated a closely-related question, i.e., the relationship 
between the ``classical" and the ``quantum" solitons, according to their terminology. 

To go beyond the mean-field approximation the method of diagonalization of the many-body Hamiltonian may be
used. The questions which are associated with the energy of the system (i.e., the dispersion relation, or the 
velocity of propagation of the waves, which is given by the slope of the dispersion relation) are attacked in 
a straightforward way from the eigenvalues of the Hamiltonian. On the other hand, extracting the density is 
much more challenging, since the eigenstates that one gets from the diagonalization of the many-body Hamiltonian 
are also eigenstates of the angular momentum and thus they are rotationally invariant. Still, the 
physically-relevant solutions are the ones which break the axial symmetry of the Hamiltonian and clearly are not 
eigenstates of the angular momentum. 

To answer the question that we are interested in, it is instructive to recall that the energy of the mean-field 
state for some fixed expectation value of the angular momentum coincides to leading order in $N$ with that of 
the corresponding ``yrast" state \cite{JKMR} (an ``yrast" state is the lowest-energy eigenstate of the Hamiltonian 
and is also an eigenstate of the angular momentum), however, the yrast state has a lower energy to subleading order 
in $N$. 

Having this in mind, we adopt the following strategy, which is based on the minimization of the energy: First 
of all, we evaluate the yrast states (diagonalizing the many-body Hamiltonian). Then, we evaluate the corresponding 
product, mean-field-like many-body state. Projecting this state on the yrast state of some given angular momentum, 
we evaluate the amplitude that corresponds to the yrast state of this specific value of the angular momentum. Using
these amplitudes, we thus construct a many-body state which is a linear superposition of yrast states. This state has 
an energy which coincides to leading order in $N$ with that of the corresponding yrast state, and of the mean-field 
state, but it has a lower energy to subleading order in $N$, even than the yrast state, provided that the effective 
interaction between the atoms is repulsive. 

In what follows below we first describe our model in Sec.\,II. In Sec.\,III we present the strategy that we 
follow for the construction of the many-body state. In Sec.\,IV we apply our method to the case of weak interactions, 
where we solve the (two-state) problem analytically. In Sec.\,V we go beyond the two-state model, presenting our 
numerical results and the finite-$N$ corrections that we are interested in. In Sec.\,VI we investigate the asymptotic 
form of the many-body state, which reduces to the well-known state of the mean-field approximation in the appropriate 
limit of a large atom number. In Sec.\,VII we discuss the experimental relevance of our study. Finally, in Sec.\,VIII 
we give a summary of the main results and our conclusions.

\section{Model and general considerations}

We assume for simplicity one-dimensional motion of bosonic atoms under periodic boundary conditions, as in a 
ring potential. This assumption is valid provided that the interaction energy is much smaller than the quantum 
of energy of the trapping potential in the transverse direction, in which case the transverse degrees of freedom 
are frozen. 

If ${\hat a}_m$ and ${\hat a}_m^{\dagger}$ are annihilation and creation operators of an atom with angular momentum 
$m \hbar$, the Hamiltonian that we consider has the form
\begin{eqnarray}
  {\hat H} = \frac {\hbar^2} {2 M R^2} \sum_{m=m_{\rm min}}^{m = m_{\rm max}} m^2 {\hat a}_m^{\dagger} {\hat a}_m 
  + \frac {U} 2 \sum_{m+n=k+l} {\hat a}_m^{\dagger} {\hat a}_n^{\dagger} a_k a_l,
\nonumber \\ \label{ham11}
\end{eqnarray}
where $m_{\rm min}$ and $m_{\rm max}$ are the lowest and the highest values of $m$ that we consider, $M$ is the atom 
mass, $R$ is the radius of the ring, and $U$ is the matrix element for elastic s-wave atom-atom collisions (assumed 
to be positive). There are thus two energy scales in the problem, namely the kinetic energy per particle $\epsilon = 
\hbar^2/(2 M R^2)$ associated with the motion of the atoms along the ring, and the interaction energy per particle, 
which for a homogeneous gas is equal to $(N-1) U/2$. It is thus convenient to introduce the dimensionless quantity 
$\gamma$ as the ratio between the interaction energy $(N-1) U$ and the kinetic energy $\epsilon$, $\gamma = (N-1) 
U/\epsilon$.

\section{Constructing the many-body state as a superposition of yrast states}

The first step in our calculation is the evaluation of the yrast states $|\Phi_{\rm ex}(L) \rangle$, which are
eigenstates of the Hamiltonian $\hat H$, of the operator of the angular momentum $\hat L$ and of the number
operator $\hat N$. As mentioned above, the Hamiltonian is axially symmetric and thus the eigenstates respect 
this symmetry, which implies that the corresponding single-particle density distribution is axially symmetric.
Indeed, if $\hat{\Psi}(\theta)$ is the destruction operator of a particle at an angle $\theta$, then the 
single-particle density is,
\begin{eqnarray}
 n(\theta) &=& \langle \Phi_{\rm ex}(L) | {\hat \Psi}^{\dagger}(\theta) {\hat \Psi}(\theta) | \Phi_{\rm ex}(L) 
 \rangle = \nonumber \\ &=& \frac 1 {2 \pi R}
  \sum_{m,m'} \langle \Phi_{\rm ex}(L) | {\hat a}_m^{\dagger} {\hat a}_{m'} | \Phi_{\rm ex}(L) \rangle 
  \, e^{i(m'-m) \theta} =
\nonumber \\ &=& \frac {N} {2 \pi R}, 
\label{nex}
\end{eqnarray}
i.e., $n(\theta)$ is spatially independent and equal to the mean density for any value of $L$, since the matrix 
elements appearing above are diagonal. 

On the other hand, the physically-relevant solutions are not axially symmetric (in general), and this is a major 
problem. Actually, this problem has a much more general aspect, namely the relationship between the mean-field 
solutions, which break the axial symmetry of the problem, and the eigenstates of the many-body Hamiltonian -- 
where one is working with eigenstates of the angular momentum, too, and as a result they respect the axial symmetry 
of the problem. Here the general strategy that we develop allows us to extract the spatially-dependent single-particle 
density distribution from the yrast, many-body eigenstates, going beyond the mean-field approximation.

It should be mentioned that a method that is often used to overcome this difficulty of breaking of the axial 
symmetry is to introduce correlation functions, for example \cite{RMP}, 
\begin{eqnarray}
n^{(2)}(\theta, \theta_0) \propto \langle \Phi_{\rm ex}(L) | {\hat \Psi}^{\dagger}(\theta) {\hat \Psi}^{\dagger}
(\theta_0) {\hat \Psi}(\theta_0) {\hat \Psi}(\theta) | \Phi_{\rm ex}(L) \rangle, 
\end{eqnarray}
whre $\theta_0$ is some reference point. While this method does indeed break the axial symmetry and allows us to 
get a qualitative answer, it cannot be trusted quantitatively. The easiest example is that of weak interactions 
(examined in detail below), where it turns out that 
\begin{eqnarray}
n^{(2)}(\theta, \theta_0) &\propto& [N (N-1)+ 2 L (N-L) \cos(\theta - \theta_0)]
\nonumber \\
                          &\propto& (1-1/N) + 2 \ell (1 - \ell) \cos(\theta - \theta_0), 
\end{eqnarray}
where $\ell = L/N$. The above expression cannot in any way be related to the density that results from the mean-field 
approximation, 
\begin{eqnarray}
 n_{\rm MF} (\theta) = \frac N {2 \pi R} (1 + 2 \sqrt {\ell (1 - \ell)} \cos \theta),
\label{mfsw}
\end{eqnarray} 
not even in the limit of large values of $N$. (In the above expression we have assumed that the arbitrary position of 
the minimum of the density is at $\theta = 0$.) Therefore, we conclude that $n^{(2)}(\theta, \theta_0)$ cannot be used 
for quantitative comparisons.

The way that we proceed is thus the following. We introduce an essentially variational many-body state, namely
\begin{eqnarray}
 |\Phi(\ell_0) \rangle = {\cal C}\sum_{L=L_{\rm min}}^{L_{\rm max}} \langle \Phi_{\rm ex}(L) | \Phi_{\rm MF} (\ell_0) 
 \rangle \, |\Phi_{\rm ex}(L) \rangle,
\label{finaleq}
\end{eqnarray}  
where ${\cal C}$ is the normalization constant. In other words, we take the inner product between the mean-field state 
with some angular momentum per atom $\ell_0 \hbar$, $|\Phi_{\rm MF}(\ell_0) \rangle$, and some yrast state with total 
angular momentum $L \hbar$, $|\Phi_{\rm ex}(L) \rangle$, to get the amplitudes $\langle \Phi_{\rm ex}(L) | \Phi_{\rm MF}
(\ell_0) \rangle$. From these amplitudes we then construct a linear superposition of eigenstates $|\Phi_{\rm ex}(L) 
\rangle$, which constitute the many-body state $|\Phi(\ell_0)\rangle$. 

This state has the following crucial features: (i) it has the desired expectation value of angular momentum, (ii) 
to leading order in $N$ it has the same energy as the mean-field state, as well as the yrast state, but it has a 
lower energy to subleading order in $N$, (iii) it gives the same single-particle density distribution as the 
mean-field state for large values of $N$, and (iv) finally it is not fragmented. 

Turning to $|\Phi_{\rm MF}(\ell_0) \rangle$, this is a product many-body state, which corresponds to the order 
parameter of the mean-field approximation; obviously we work in the same basis of single-particle  states, with 
$m_{\rm min} \le m \le m_{\rm max}$, as in the method of diagonalization, 
\begin{eqnarray}
 |\Phi_{\rm MF} (\ell_0) \rangle &=& \frac 1 {\sqrt{N!}} \left( \sum_{m=m_{\rm min}}^{m_{\rm max}} 
 c_m {\hat a}_m^{\dagger} \right)^N |0 \rangle,
\label{gmfstt}
\end{eqnarray}
where $|0\rangle$ denotes the vacuum; also $c_m$ are real, variational parameters, which we evaluate by minimizing 
the corresponding expectation value of the energy
\begin{eqnarray}
 E_{\rm MF} (\ell_0) = N \frac {\hbar^2} {2 M R^2} \sum_{m=m_{\rm min}}^{m_{\rm max}} m^2 c_m^2 + 
 \nonumber \\
 + \frac 1 2 N (N-1) U \int \left| \sum_{m=m_{\rm min}}^{m_{\rm max}} c_m \phi_m \right|^4 \, d \theta. 
\label{gmfsttt}
\end{eqnarray}
Here $\phi_m = e^{i m \theta}/\sqrt{2 \pi}$ are the single-particle eigenstates of the ring potential with 
an eigenvalue of the angular momentum equal to $m \hbar$ and an eigenenergy $\epsilon_m = m^2 \epsilon$. The 
normalization imposes the constraint $\sum c_m^2 = 1$, while there is the additional constraint that comes 
from the expectation value of the angular momentum being $\ell_0 = L_0/N$, $\sum m c_m^2 = \ell_0$.

\section{Results of our method for weak interactions}

We start with the limit of weak interactions, where we can solve this problem analytically, and then proceed 
with the more general problem of stronger interactions. 

To define the limit of ``weak"/``strong" interactions let us introduce quite generally the ratio between the 
interaction energy per particle of the homogeneous gas $(N-1) U/2$ and the kinetic energy $\epsilon_m = m^2 
\hbar^2/(2 M R^2)$, which is $\gamma/(2 m^2)$. For some given value of $\gamma$, setting $\gamma/(2 m^2) \sim 1$, 
the maximum value of $|m|$ has to be (much) larger than $\sqrt{\gamma/2}$ in order to achieve convergence. In terms 
of length scales, the parameter $\sqrt{\gamma}$ gives the ratio between the radius of the ring $R$ and the coherence 
length $\xi$ (ignoring terms of order unity). The limit $\gamma \ll 1$ defines the regime of ``weak" interactions, 
where $\xi \gg R$, while for $1 \ll \gamma \ll N^2$, then $R/N \ll \xi \ll R$, which is the ``Thomas-Fermi" limit. 
When $\gamma$ becomes of order $N^2$, then the system approaches the Tonks-Girardeau limit, where the coherence 
length $\xi$ becomes comparable to the inter-particle spacing $R/N$ and correlations play a crucial role.  

When $\gamma \ll 1$ one may work with the single-particle states $\phi_0$ and $\phi_1$ only. In this case 
$|\Phi_{\rm ex}(L) \rangle$ has the very simple form (because of the two constraints)
\begin{eqnarray}
 |\Phi_{\rm ex}(L) \rangle = |0^{N-L}, 1^L \rangle.
\end{eqnarray}
In the above notation the state $\phi_0$ has $N-L$ atoms, and the state $\phi_1$ has $L$ atoms. 

The mean-field, many-body state is
\begin{eqnarray}
 |\Phi_{\rm MF} (\ell_0) \rangle &=& \frac 1 {\sqrt{N!}} (c_0 {\hat a}_0^{\dagger} + c_1 {\hat a}_1^{\dagger})^N
 |0 \rangle
 \nonumber \\
 &=& \sum_{L=0}^N \frac {\sqrt{N!}} {(N-L)! L!}
 \, c_0^{N-L} c_1^{L} ({\hat a}_0^{\dagger})^{N-L} ({\hat a}_1^{\dagger})^{L} |0 \rangle 
 \nonumber \\
 &=& \sum_{L=0}^N \frac {\sqrt{N!}} {\sqrt{(N-L)! L!}} \, c_0^{N-L} c_1^{L} |0^{N-L}, 1^L \rangle
  \nonumber \\
 &\equiv& \sum_{L=0}^N d_L(\ell_0) |0^{N-L}, 1^L \rangle.
\label{mfstt}
\end{eqnarray}
The actual value of $c_0^2$ is $1-\ell_0$, while $c_1^2 = \ell_0$. 

The amplitudes of the above state of Eq.\,(\ref{mfstt}) have the interesting feature that 
\begin{eqnarray}
  |d_L (\ell_0)|^2 &\equiv& \frac {N!} {(N-L)! L!} \, c_0^{2(N-L)} c_1^{2L} \nonumber \\
  &\approx& \frac {e^{-(L-N \ell_0)^2/[2 N \ell_0 (1-\ell_0)]}} {\sqrt{2 \pi N \ell_0 (1-\ell_0)}}
   \left[ 1 + {\cal O} \left( \frac 1 {\sqrt N} \right) \right],
   \nonumber \\
\label{amplit}
\end{eqnarray}
where the approximate expression holds for large $N$ and $\ell_0 (1-\ell_0)$ not close to zero. Therefore, 
$|d_L|^2$ is a Gaussian, with its peak at $L_0 = N \ell_0$ (scaling as $N$) and a width which is of order 
$\sqrt{N}$, which becomes a delta function in the limit of large $N$. The above observations are generic 
features and not specific to the two-state model and thus are central in the analysis that follows below.

In the final step of our calculation we evaluate $|\Phi(\ell_0) \rangle$, which is, as described earlier,
\begin{eqnarray}
 |\Phi(\ell_0) \rangle = {\cal C} \sum_{L=0}^{N} d_L (\ell_0) |0^{N-L}, 1^L \rangle.
 \label{finaleqq}
\end{eqnarray}  
In the two-state approximation, $|\Phi(\ell_0) \rangle$ coincides with $|\Phi_{\rm MF} (\ell_0) \rangle$. 
However, we stress that this is not a general result, as we explain below.
 
The single-particle density matrix of $|\Phi(\ell_0) \rangle$, $\rho_{ij} = \langle \Phi(\ell_0) |a_i^{\dagger} 
a_j| \Phi(\ell_0) \rangle$ (with $i,j = 0, 1$), is 
\[ \rho = N \left( \begin{array}{cc}
c_0^2 & c_0 c_1 \\
c_0 c_1 & c_1^2 \end{array} \right).\] 
The two eigenvalues are $\lambda = 0$ and $\lambda = 1$ (the determinant of the above matrix vanishes and thus 
one of the eigenvalues has to vanish). The state $|\Phi(\ell_0) \rangle$ is not fragmented, as one expects. The 
eigenvector that corresponds to $\lambda = 1$ is the expected one, namely $\psi = c_0 \phi_0 + c_1 \phi_1$. The 
single-particle density distribution in $|\Phi(\ell_0) \rangle$ is given by
\begin{eqnarray}
 n(\theta) &=& \langle \Phi(\ell_0) | {\hat \Psi}^{\dagger}(\theta) {\hat \Psi}(\theta) | \Phi(\ell_0) \rangle = 
\nonumber \\ &=& \frac N {2 \pi R} [1 + 2 |c_0| |c_1| \cos (\theta - \theta_0)]
\nonumber \\ &=& \frac N {2 \pi R} [1 + 2 \sqrt{\ell_0 (1-\ell_0)} \cos (\theta-\theta_0)],
\label{nexxx}
\end{eqnarray}
where $\theta_0$ is the relative phase between $c_0$ and $c_1$. This phase is arbitrary reflecting the rotational 
invariance of the Hamiltonian; assuming that $c_0$ and $c_1$ are real, it is equal either to zero, or $\pi$. 
However, this degeneracy is lifted when the symmetry is broken (see more in the following paragraph). The result 
of Eq.\,(\ref{nexxx}) coincides with that of Eq.\,(\ref{mfsw}), and therefore in the two-state model the density 
that one gets from our many-body state $|\Phi(\ell_0) \rangle$ is the same as the one derived from the Jacobi 
(sinusoidal) solutions of the mean-field Gross-Pitaevskii approximation.

Furthermore, the expectation value of the energy ${\cal E}$ in the state $|\Phi(\ell_0) \rangle$ is
\begin{eqnarray}
  {\cal E} - \frac U 2 N (N-1) = \frac {\hbar^2 L_0} {2 M R^2} + U L_0 (N-L_0) (1-1/N),
 \nonumber \\ \label{eqq1}
\end{eqnarray}
where again the phase $\theta_0$ does not appear in the energy due to the assumed axial symmetry of the Hamiltonian. 
When a ``weak" symmetry breaking potential $\Delta V = V_0 \cos \theta$ is added to the Hamiltonian, then 
\begin{eqnarray}
{\cal E} - U N (N-1)/2 &=& \frac {\hbar^2 L_0} {2 M R^2} + U L_0 (N-L_0) (1-1/N) +
\nonumber \\
&+& \cos \theta_0 N V_0 \sqrt{\ell_0 (1-\ell_0)}/2. 
\label{eqq2}
\end{eqnarray}
Here we see an explicit dependence of the energy on $\theta_0$. Obviously $\theta_0$ has to take the value $\pi$ 
in order for the energy to be minimized. As a result, the single-particle density is $n(\theta) = N (1 - 2 
\sqrt{\ell (1 - \ell)} \cos \theta)/(2 \pi R)$, and thus the minimum of the density is at $\theta = 0$, where the 
value of the potential is maximum.

In Eqs.\,(\ref{eqq1}) and (\ref{eqq2}) above we observe that to leading order in $N$ the interaction energy agrees 
with that of the eigenstates $|0^{N-L_0}, 1^{L_0} \rangle$, ${\cal E}_{\rm ex}$, which is
\begin{eqnarray}
 {\cal E}_{\rm ex} - \frac {U} 2 N (N-1) = \frac {\hbar^2 L_0} {2 M R^2} + U L_0 (N-L_0),
\label{diag22}
\end{eqnarray}
but is lower to subleading order. This result is due to the fact that the dispersion relation has a negative 
curvature and the fact that $|\Phi(\ell_0) \rangle$ samples other states -- of order $\sqrt N$ -- around the 
``pure" state $|\Phi_{\rm ex} (L_0) \rangle = |0^{N-L_0}, 1^{L_0} \rangle$. Actually, this lowering of the energy 
to subleading order in $N$ of the state that breaks the axial symmetry reflects precisely this fact, namely that 
the curvature of the dispersion relation is negative (provided that the effective interaction is repulsive). 

Indeed, quite generally, a distribution $P(L) \propto e^{-(L-L_0)^2/N}/\sqrt{N}$ [see Eq.\,(\ref{amplit})] gives 
an average of $U L (N-L)$ which differs from $U L_0 (N-L_0)$ [see Eq.\,(\ref{diag22})] that is proportional to 
$-U N$, as in Eq.\,(\ref{eqq1}), 
\begin{eqnarray}
U \langle L (N-L) \rangle - U L_0 (N-L_0) &\propto& - U N \, {\rm erf}(L_0/\sqrt N) 
\nonumber \\ &\approx& - U N,
\label{arg}
\end{eqnarray}
where ${\rm erf}(x)$ is the error function.

\section{Beyond the two-state model: Finite-$N$ corrections and numerical results}

The two-state model discussed above has the advantage that all the calculations may be performed analytically.
In addition, it provides an accurate description of $|\Phi(\ell_0) \rangle$ in the limit of weak interactions, 
$\gamma \ll 1$. The results of the previous section demonstrate that in this limit the state $|\Phi(\ell_0) 
\rangle$ -- even in a system with a finite number of atoms $N$ -- essentially coincides with the one of the 
mean-field approximation (which, in the limit of weak interactions, gives a sinusoidal density distribution).

On the other hand, as we also saw earlier, the very drastic restriction to the states $\phi_0$ and $\phi_1$ 
(only) forces the yrast states to have the trivial form $|0^{N-L}, 1^L \rangle$ and therefore all the angular 
momentum is carried by the single-particle state $\phi_1$. It is thus necessary to work in a more extended space. 
In the interval $0 \le \ell \le 1$ and for relatively stronger interactions in addition to $\phi_0$ and $\phi_1$, 
$\phi_2$ and $\phi_{-1}$ have the most significant contribution, while if one wants to go even further $\phi_3$ 
and $\phi_{-2}$ have to be included, too, and so on, as Bloch's theorem implies \cite{FB}. 

Before we turn to a higher truncation, it is instructive to examine the case with $\phi_0$ and $\phi_1$, and $\phi_2$ 
only. In this subspace the yrast states have the form
\begin{eqnarray}
  |\Phi_{\rm ex}(L) \rangle = \sum_m (-1)^m b_m |0^{N-L+m}, 1^{L-2m}, 2^{m} \rangle,
  \label{yrst}
\end{eqnarray}
where the amplitudes $b_m$ are Gaussian distributed \cite{JKMR}. The corresponding mean-field, product many-body
state has the form 
\begin{eqnarray}
 |\Phi_{\rm MF} (\ell_0) \rangle = \frac 1 {\sqrt{N!}} (c_0 {\hat a}_0^{\dagger} + c_1 {\hat a}_1^{\dagger} 
 + c_2 {\hat a}_2^{\dagger})^N |0 \rangle
 \nonumber \\
 = \sum_{k=0}^N \sum_{m=0}^k  
 \frac {\sqrt{N!} \, c_0^{N-k} c_1^{k-m} c_2^m} {\sqrt{(N-k)! (k-m)! m!}} |0^{N-k}, 1^{k-m}, 2^m \rangle
 \nonumber \\
 \equiv \sum_{k=0}^N \sum_{m=0}^k d_{m,k} |0^{N-k}, 1^{k-m}, 2^m \rangle. 
\label{mfsttt}
\end{eqnarray} 
Taking the inner product between the states $|\Phi_{\rm ex}(L) \rangle$ and $|\Phi_{\rm MF} (\ell_0) \rangle$ 
forces the index $``k"$ to be equal to $L-m$ and therefore the only non-zero amplitudes are $d_{m, L-m}$, which 
are also Gaussian distributed. When more single-particle states are included in the calculation the general 
picture is the same, with the only difference being that one has a multi-dimensional space; if $r$ is the number 
of single-particle states, the dimensionality is $r-2$, because of the two constraints. In each direction in this 
space one still obtains Gaussian distributions and on a qualitative level the final result is essentially the same.

\begin{figure}
\includegraphics[width=7cm,height=5.cm]{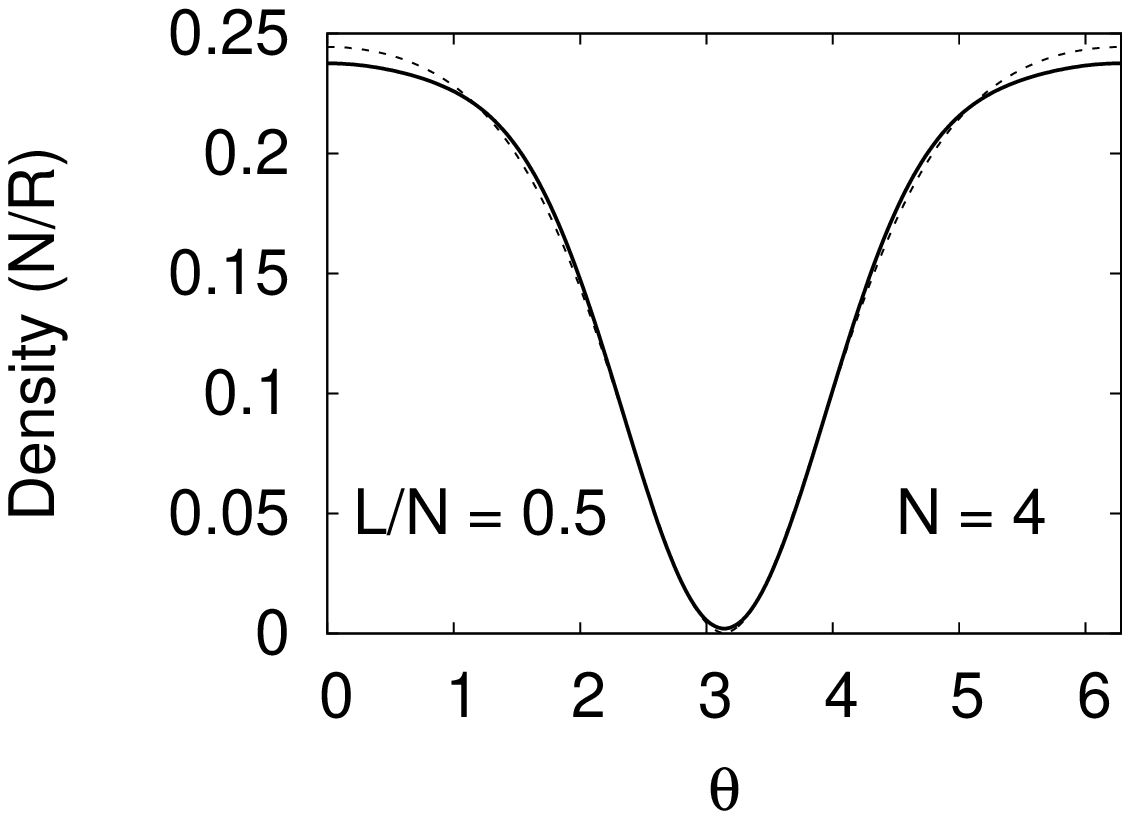}
\includegraphics[width=7cm,height=5.cm]{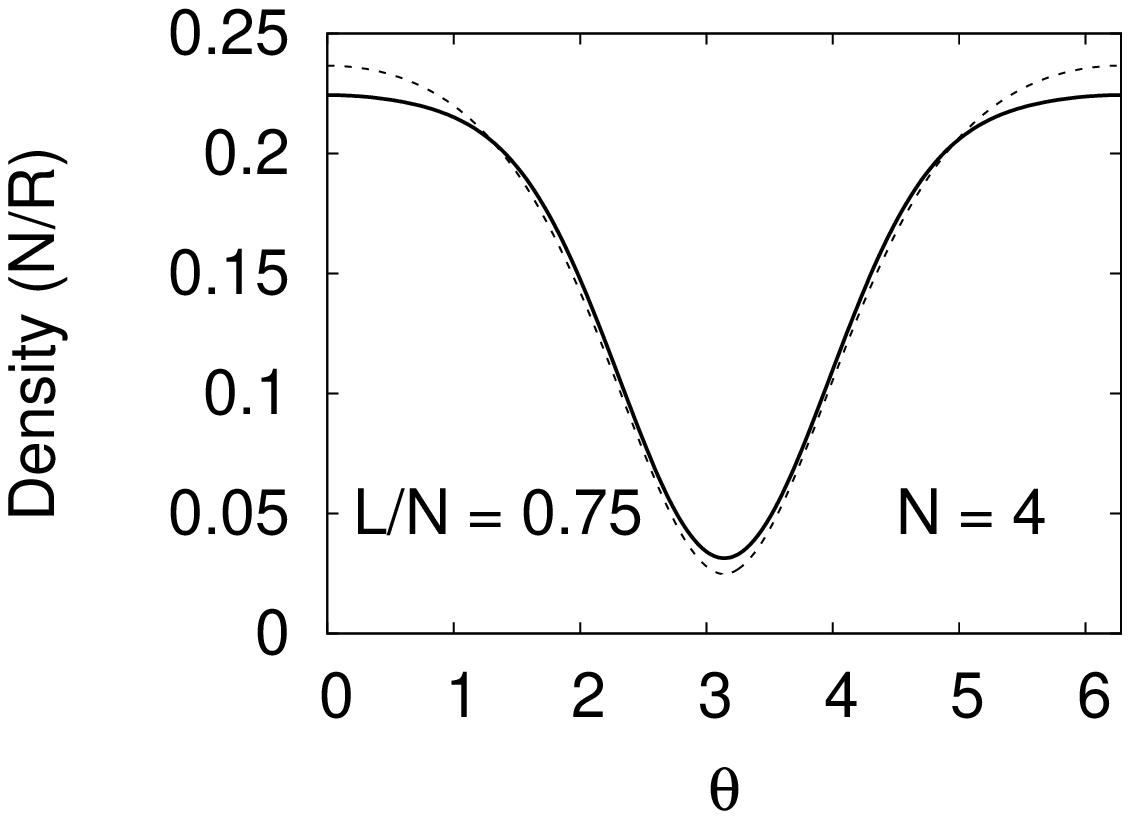}
\includegraphics[width=7cm,height=5.cm]{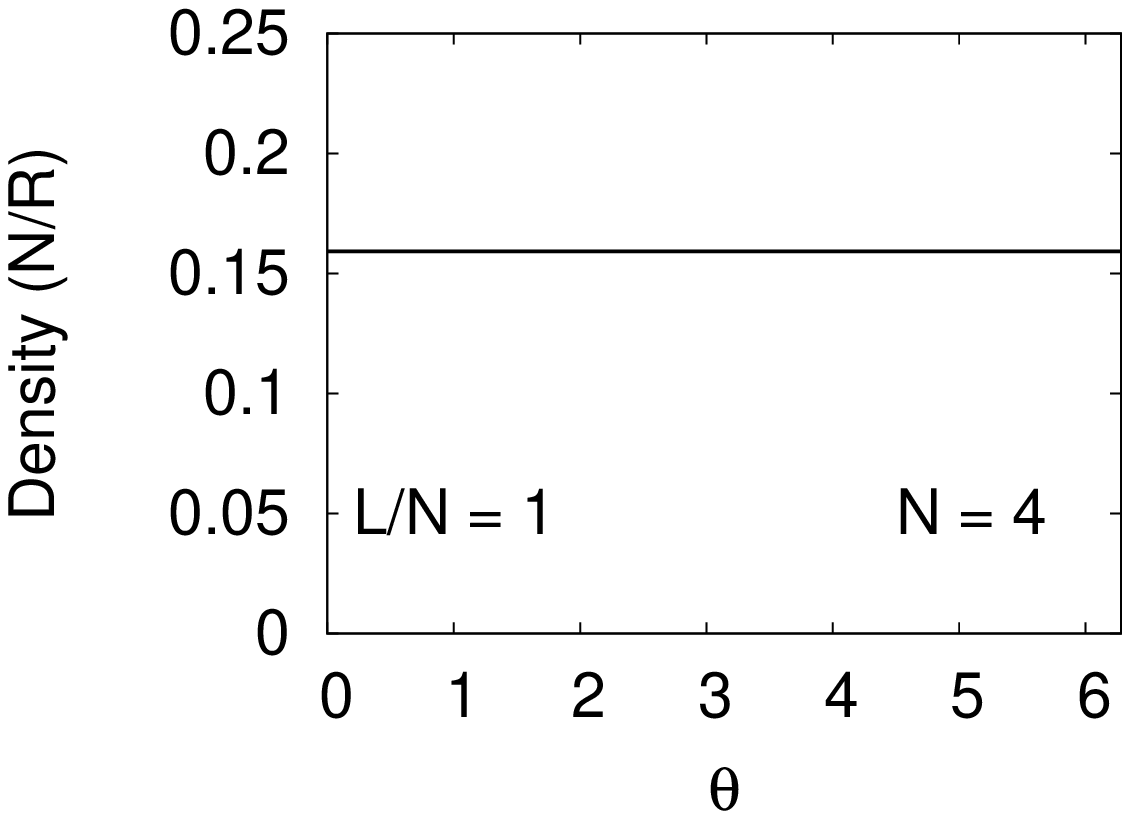}
\vskip1.5pc
\caption{Solid lines: The single-particle density distribution $n(\theta)$ corresponding to $|\Phi(\ell) \rangle$ 
of a finite system of atoms, within the space of states with $m = -2, \dots, 3$, for $N = 4$ atoms, $\gamma 
= (N-1) U/\epsilon = 0.9$, and $\ell = L/N = 0.50, 0.75$, and 1.00, from top to bottom. Dashed lines: 
The single-particle density distribution $n_{\rm MF}(\theta)$, corresponding to $|\Phi_{\rm MF} (\ell) \rangle$, 
derived within the mean-field approximation, via the minimization of the energy, for the same set of parameters.}
\end{figure}

\begin{figure}
\includegraphics[width=7cm,height=5.cm]{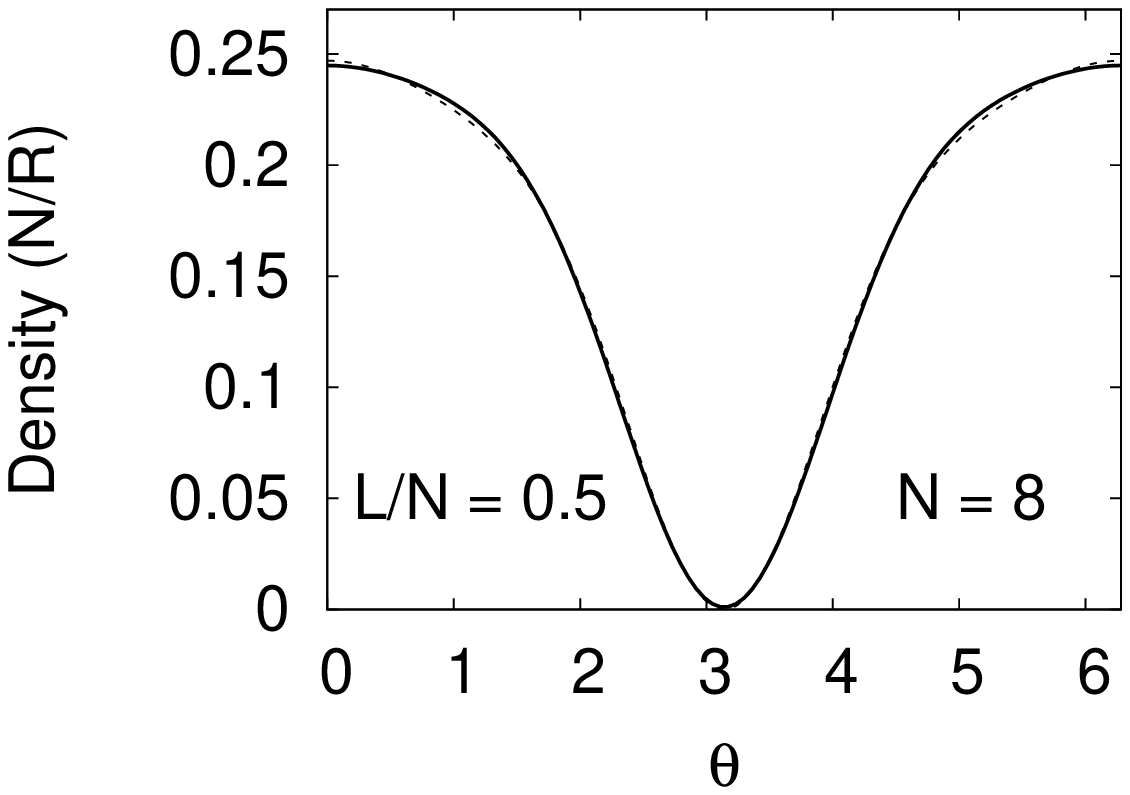}
\includegraphics[width=7cm,height=5.cm]{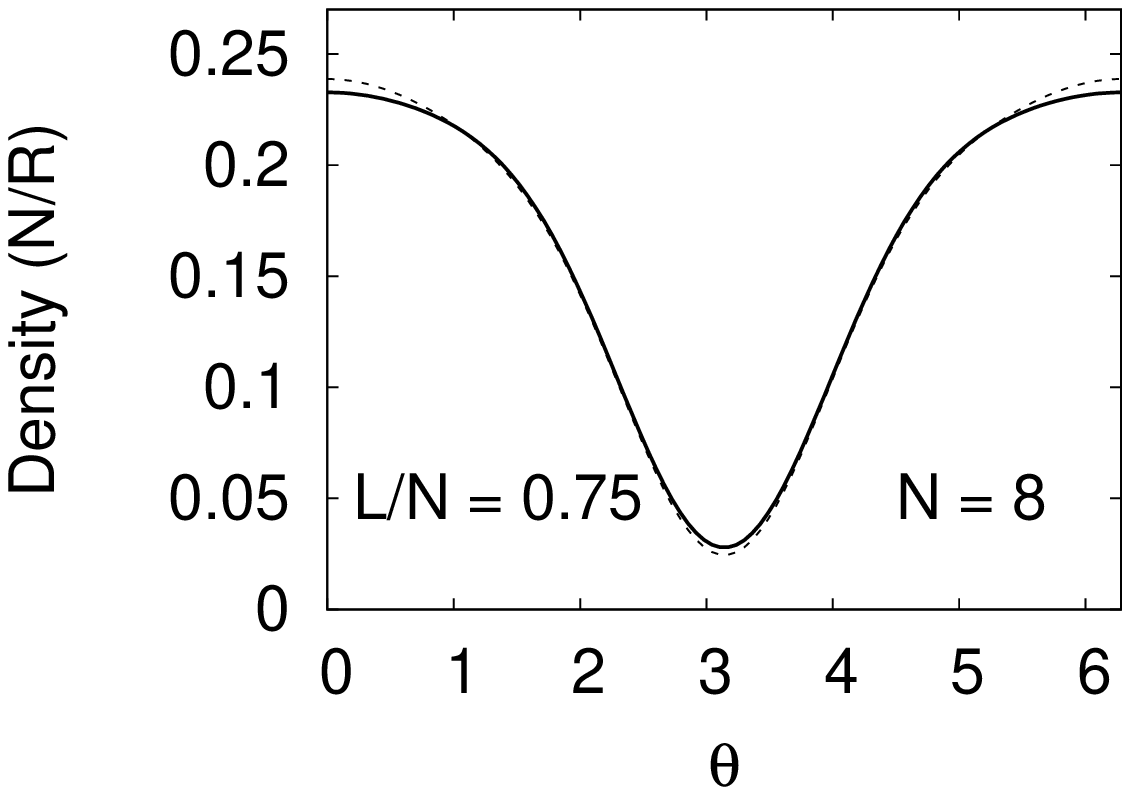}
\includegraphics[width=7cm,height=5.cm]{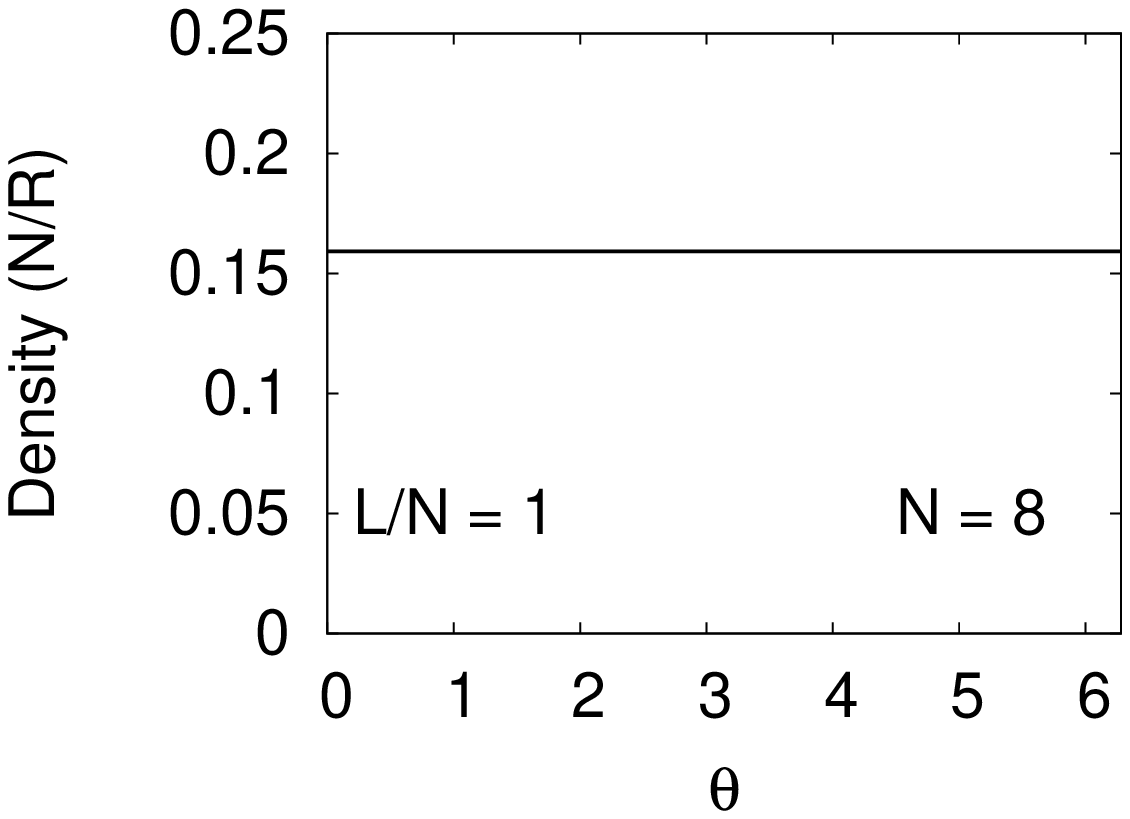}
\vskip1.5pc
\caption{Same as Fig.\,1, with $N = 8$ and $\gamma = 0.9$. The difference between the two curves is hardly visible.}
\end{figure}

\begin{figure}
\includegraphics[width=7cm,height=5cm]{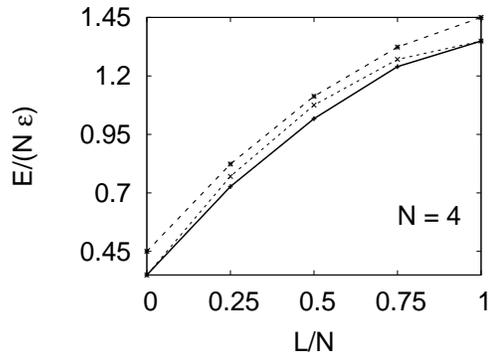}
\vskip1.5pc
\caption{The energy per particle as function of the angular momentum per particle for $N = 4$ atoms, $\gamma = 
(N-1) U/\epsilon = 0.9$, and $\ell = L/N = 0.50, 0.75$, and 1.00, within the space of states with $m = -2, \dots, 
3$. The lowest data corresponds to the expectation value of the energy of $|\Phi(\ell) \rangle$, the middle to 
the eigenvalues of the yrast states $|\Phi_{\rm ex} (L) \rangle$, and the top to the energy of the mean-field
state $|\Phi_{\rm MF}(\ell) \rangle$. We stress that in the middle curve $L$ on the x-axis is the eigenvalue of 
${\hat L}$, while in the other two curves $L$ is the expectation value of $\hat L$.}
\end{figure}

We proceed now with the calculation within the states $\phi_{-2}, \phi_{-1}, \phi_0, \phi_1, \phi_2$ and $\phi_3$. 
We stress that $\gamma$ has to be sufficiently large in order to have a significant occupancy of the states other
than the ones with $m=0$ and $m=1$, since otherwise one goes back to the two-state model described earlier. In addition 
we have to make sure that we have reached convergence with respect to the number of single-particle states that are 
considered and finally we have to ensure that Bloch's theorem \cite{FB} is not violated. As described also above, we 
first evaluate the yrast states $| \Phi_{\rm ex}(L) \rangle$ for $-2N \le L \le 3N$, diagonalizing the many-body 
Hamiltonian. We also evaluate $|\Phi_{\rm MF} (\ell_0) \rangle$
\begin{eqnarray}
 |\Phi_{\rm MF} (\ell_0) \rangle = \frac 1 {\sqrt{N!}} \left( \sum_{m=-2}^{m=3} c_{m} {\hat a}_{m}^{\dagger} \right)^N 
 |0 \rangle,
\end{eqnarray} 
where $c_{-2}, c_{-1}, c_0, c_1, c_2$ and $c_3$ are evaluated from the constraint minimization of the energy. In the 
final step we evaluate $|\Phi(\ell_0) \rangle$ using Eq.\,(\ref{finaleq}), which then gives us the single-particle 
density distribution (or any other observable). 

In Figs.\,1 and 2 we plot the corrections of the density in a finite system of atoms for various values of the angular
momentum. The solid curves show the single-particle density $n(\theta)$ that corresponds to $|\Phi (\ell_0)\rangle$, 
while the dashed ones show the density $n_{\rm MF}(\theta)$ that corresponds to the mean-field state 
$|\Phi_{\rm MF}(\ell_0) \rangle$. In Fig.\,1 $N = 4$, and in Fig.\,2 $N = 8$, while $\gamma = 0.9$ in both of them. 
Also $\ell_0$ takes the three values $\ell_0 = 0.5, 0.75$, and 1. According to Bloch's theorem \cite{FB} for any $0 \le 
\ell_0 \le 1$, the density distribution is the same also for $\ell_0' = 1 - \ell_0$, as well as for $\ell_0' = \ell_0 
+ \kappa$, where $\kappa$ is an integer. In other words, the three graphs for $\ell_0 = 1 \, (L=4), \ell_0 = 3/4 \, 
(L=3)$, and $\ell_0 = 1/2 \, (L=2)$ shown in Fig.\,1 cover all the possible values of the angular momentum of the 
whole spectrum (for $N=4$).

For $\ell_0 = 1$ both $n_{\rm MF}(\theta)$, as well as $n(\theta)$ are constant, even for small values of $N$. We stress, 
however, that the two states are different, since $|\Phi(\ell_0) \rangle$ coincides with the yrast state for $L = N = 4$. 
For example, even within the space with $m = 0, 1$, and 2, 
\begin{eqnarray}
|\Phi_{\rm MF} (\ell_0=1) \rangle = |0^0, 1^4, 2^0 \rangle, 
\end{eqnarray}
while 
\begin{eqnarray}
|\Phi (\ell_0 = 1) \rangle &=& A_1 |0^0, 1^4, 2^0 \rangle + A_2 |0^1, 1^2, 2^1 \rangle  
\nonumber \\
&+& A_3 |0^2, 1^0, 2^2 \rangle, 
\nonumber \\
\end{eqnarray}
where $A_i$ are constants. As a result, the energy of the two states is also different, (see Fig.\,3, which shows the 
dispersion relation, that we discuss in the next section).

Returning to the density shown in Figs.\,1 and 2, while for $\ell_0 = 1$ the situation is not interesting (at least 
with regards to the density), for $\ell_0 = 1/2$ and $\ell_0 = 3/4$ there are deviations (between the dashed and the
solid curves). For $\ell_0 = 1/2$ within the mean-field approximation, the ``dark" solitary wave forms and $n_{\rm MF}
(\theta)$ has a node. As seen from Fig.\,1, $n(\theta)$ still has a node. The most significant deviations appear at 
the maxima of the density. Our state of lower energy flattens out within a larger interval compared to the mean-field 
density. These deviations (almost) disappear for $N = 8$, as shown in Fig.\,2, where $\gamma$ is still 0.9 (we discuss 
the asymptotic behavior of $|\Phi(\ell) \rangle$ in the following section). Finally, for $\ell_0 = 3/4$, there is still 
a significant deviation between $n(\theta)$ and $n_{\rm MF}(\theta)$, with roughly the same characteristics as the case 
$\ell_0 = 1/2$. 

Although the difference between the density is small (due to the relatively small value of $\gamma$, which makes
$|c_0|$ and $|c_1|$ to be much larger than all the other coefficients -- for example, for $\ell = 1/2$, $|c_{0}| = 
|c_1|$ are roughly 7 times larger than $|c_{-1}| = |c_2|$), still these results are generic. Increasing $\gamma$
will give more pronounced differences, however the problem becomes more demanding computationally, since we also
need to make sure that convergence with respect to the single-particle states that we have considered has been 
achieved. 
  
\section{Asymptotic limit of the many-body state}

It is crucial to confirm that the observables from the state that we have introduced coincide with the ones of the 
mean-field state in the appropriate limit of large $N$. The relevant limit is the one where $N$ increases, with the 
ratio between the interaction energy and the kinetic energy kept fixed (which in our notation is $\gamma$).

As seen already in the previous section, the density indeed approaches that of the mean-field state. Turning to the
energy, as argued in Sec.\,IV, in the above limit, the dominant amplitude in $|\Phi (\ell_0) \rangle$ is the one that 
corresponds to $L_0 = N \ell_0$, which is the yrast state $|\Phi_{\rm ex} (L_0) \rangle$ with this value of $L = L_0$. 
As shown in Ref.\,\cite{JKMR} $|\Phi_{\rm ex} (L_0) \rangle$ has the same energy to leading order in $N$ as the 
mean-field state $|\Phi_{\rm MF} (\ell_0) \rangle$, and the same result is true for $|\Phi(\ell_0) \rangle$. 
Therefore, all these three states have the same energy to leading order in $N$. 

Still, we stress that there is a clear hierarchy of the energies of the three states to subleading order in $N$: 
$|\Phi_{\rm MF} (\ell_0) \rangle$ has the highest energy, $|\Phi_{\rm ex} (L_0) \rangle$ has a lower energy, and 
$|\Phi(\ell_0) \rangle$ has the lowest. The first inequality has been analysed in Ref.\,\cite{JKMR}, while the 
second is due to the simple reason stated earlier [see Eq.\,(\ref{arg})].

In Fig.\,3 we plot the dispersion relation, which is evaluated within the mean-field approximation, the energy of 
the corresponding eigenstates of the many-body Hamiltonian, and the energy of $|\Phi(\ell_0) \rangle$ that we have 
evaluated. In the middle curve $L$ on the x-axis is the eigenvalue of ${\hat L}$, while in the other two curves $L$ 
is the expectation value of $\hat L$. These results provide full support of the arguments we made about the hierarchy 
of the energies. We should also mention that according to Bloch's theorem \cite{FB}, the total energy spectrum (i.e., 
for higher values of $\ell$) is the one shown in Fig.\,3, on top of a parabola.

Turning to the question of fragmentation, the single-particle density matrix of $|\Phi(\ell_0) \rangle$, 
$\rho_{ij} = \langle \Phi (\ell_0) |{\hat a}_i^{\dagger} {\hat a}_j| \Phi (\ell_0) \rangle = c_i c_j$. This result 
generalizes the one we found earlier for the case of two modes only. This matrix has one eigenvalue which is equal 
to unity, while all the other eigenvalues vanish and indeed $|\Phi (\ell_0) \rangle$ is not fragmented. The 
eigenvector that corresponds to the nonzero eigenvalue is the expected one, i.e., $\sum c_m \phi_m$.

\section{Experimental relevance}

In order to make contact with experiment, the first question is the extent that under typical conditions the
motion of the atoms is quasi-one-dimensional, as we have assumed here. If we consider the experiment of 
Ref.\,\cite{hysteresis} as an example, the system is far from this limit. In this experiment, where $N \approx 
4 \times 10^5$ $^{23}{\rm Na}$ atoms were used, their chemical potential $\mu/\hbar \approx 2 \pi \times 1.7$ 
kHz, was much larger than the frequencies of the (annular-like) trapping potential, $\omega_1 \approx 472$ Hz 
and $\omega_2 \approx 188$ Hz. The dimensionless quantity that describes the transition to the one-dimensional 
limit may also be expressed in terms of $N a/R$, which has to be $\ll 1$ in this limit. Here $R \approx 19.5$ 
$\mu {\rm m}$ is the mean radius of the torus/annulus and $a \approx 28$ \AA \, is the s-wave scattering length 
for atom-atom collisions. Since $N a/r$ is $\approx 5 \times 10^2$, the system is not in the limit of 
quasi-one-dimensional motion. 

In addition, the dimensionless parameter $\gamma = 2 N a R/S$, where $S$ is the cross section of the torus/annulus, 
for the parameters of this experiment is on the order of 1500 \cite{Alex}. As a result, this experiment is in the 
Thomas-Fermi regime (see Sec.\,IV), where $1 \ll \gamma \ll N^2$, and the solitary waves will resemble the 
well-known ones of an infinite system. 

According to the results of our study three conditions have to be satisfied in order for the corrections that 
we have predicted to be substantial. First of all, $N$ should not exceed $\sim 10$, since otherwise the corrections 
will be suppressed. Also the interaction energy has to be sufficiently strong, i.e., $\gamma$ should be at least 
of order unity, since otherwise the system is in the limit of weak interactions, where the deviations we have 
found are also suppressed. Finally, $N a/R$ has to be $\ll 1$.

If one simply reduces $N$ under the conditions of the experiment of Ref.\,\cite{hysteresis}, by e.g., a factor 
of order $10^5$, so that $N$ will become equal to 4, then $N a/R \sim 5 \times 10^{-3}$, i.e., indeed it will 
be $\ll 1$, however $\gamma \sim 1.5 \times 10^{-2}$, i.e., $\gamma$ will also become $\ll 1$, which will bring 
the system to the limit of weak interactions. Therefore, in addition to reducing $N$ by a large factor, one should 
also decrease $S \propto 1/\sqrt{\omega_1 \omega_2}$ (by, e.g., increasing the trapping frequencies in the transverse 
direction), in order to make $\gamma \sim 1$.

\section{Summary and conclusions}

To summarize, the general problem that we have investigated is the relationship between the mean-field
approximation and the method of diagonalization of the many-body Hamiltonian, in connection with the 
spontaneous breaking of the symmetry of the Hamiltonian (assumed to be axially symmetric). While the 
mean-field states break the symmetry of the Hamiltonian by construction, the eigenstates of the 
Hamiltonian respect this symmetry, thus giving rise to a single-particle density distribution which 
is always axially symmetric. Still, in a real system the axial symmetry is broken, even under very weak 
symmetry-breaking mechanisms (if one is interested in the solutions which do not break the symmetry,
these are the eigenstates of the Hamiltonian.) 

One of the main goals of the present study is to investigate how one breaks the symmetry, going also 
beyond the mean-field approximation. To achieve this goal, we make use of the mean-field approximation 
and then construct a linear superposition of the eigenstates of the many-body Hamiltonian. This state 
breaks the symmetry and is not a product state, i.e., it goes beyond the mean-field approximation. We 
stress that the method that we have developed is general and may be applied to other problems, as well.  

Another main result of our study is the actual problem where we have applied this method, namely the 
finite-$N$ corrections of the well-known solitary-wave solutions which result within the (one-dimensional) 
nonlinear Schr\"odinger equation in a finite ring. Interestingly, the state that we have used is one of 
``minimum-uncertainty" and in a sense it is the mostly ``classical". According to our results, for low 
interaction strengths, where the two-state model is a good approximation, the finite-$N$ corrections are 
negligible and one gets back to the ordinary Jacobi solutions of the nonlinear Schr\"odinger equation (which 
are sinusoidal in this limit). For larger interaction strengths and/or small atom numbers, where more than 
two single-particle states need to be considered, these corrections become non-negligible and there are 
significant deviations between our many-body state and the mean-field state. 

While we have not proven that the many-body state that we have constructed provides the absolute minimum of the 
energy, what we do know is that it has the same energy to leading order in $N$, and a lower energy to subleading
order in $N$, as compared the mean-field state, and the corresponding eigenstate of the Hamiltonian. In other
words, this many-body state provides a lower bound of the energy. One subtle point in these arguments is of course
that in this comparison the yrast state is an eigenstate of the operator of the angular momentum $\hat L$, and
thus $``L"$ is the corresponding eigenvalue. Within the other two states, since they break the symmetry, $``L"$ 
is the expectation value of $\hat L$. 

When the angular momentum is an integer multiple of $\hbar$ within the mean-field approximation the density of the
cloud is predicted to be homogeneous. According to our results, the single-particle density distribution remains 
homogeneous even for a small atom number. When the angular momentum per particle is equal to a half integer within 
the mean-field approximation the ``dark" solitary wave forms, which has a node in its density. According to our 
analysis in a system with a small atom number the single-particle density distribution still has a node, with the 
main effect of the finiteness appearing at the ``edges" of the wave. Interestingly, a ``universal" feature of the 
dark solitary wave is its velocity of propagation, which turns out to be $\hbar/(2MR)$ (for any interaction strength, 
or any atom number $N$), as we have found numerically, essentially due to Bloch's theorem \cite{FB}. Last but not 
least, deviations in the density between the two states are also present in the intermediate values of the angular 
momentum (between 0.5 and 1). These corrections vanish in the appropriate limit of large $N$. 

Our suggested many-body state is not a ``solitary-wave state" in the strict sense of a travelling-wave solution and
for this reason we have not used this terminology here. The dynamics of this many-body wavefunction is a problem 
that we are currently working on and will be analysed in a future study.

The interaction strengths that we have considered keep us away from the correlated, Tonks-Girardeau limit. It would 
be interesting to try to push this calculation to this regime \cite{tgw}. In the spirit of density functional theory, 
one may develop a mean-field description \cite{Kol} and then use the present approach, which may still provide an 
accurate description of the system. It would be interesting to study this problem and get some quantitative answers, 
which is actually what we plan to do in the future. 

\acknowledgements

We thank Andy Jackson and Stephanie Reimann for useful discussions.

\end{document}